\documentclass[reprint, amsmath, amssymb, aps, prl, longbibliography]{revtex4-2}

\usepackage{graphicx}
\usepackage{dcolumn}
\usepackage{bm}
\usepackage{hyperref}
\usepackage{scrhack}
\usepackage{float}
\usepackage{makecell}
\usepackage{booktabs}
\usepackage{boldline}

\begin{document}

\title{Correlated Matter Induced Biases in Long-Baseline Neutrino Oscillation Measurements}

\author{Tia Pandit}
\affiliation{Department of Physics, Kirori Mal College, University of Delhi, Delhi-110007, India}

\author{Bipin Singh Koranga}
\affiliation{Department of Physics, Kirori Mal College, University of Delhi, Delhi-110007, India}

\date{\today}
\begin{abstract}
We demonstrate that treating Earth matter effects via a constant-density approximation introduces a fundamental systematic error in long-baseline neutrino oscillation analyses. Using exact numerical propagation through realistic PREM profiles, we show that matter-profile mismodeling does not merely affect the $\nu_{\mu}\rightarrow\nu_{e}$ appearance probability, but generates correlated biases across the $\nu_{\mu}\rightarrow\nu_{\tau}$ and $\nu_{\mu}\rightarrow\nu_{\mu}$ channels as dictated by PMNS unitarity. Our stochastic analysis reveals that the $\nu_{\mu}\rightarrow\nu_{\tau}$ channel is the most volatile carrier of the geophysical systematic. Across varying correlation lengths at baselines like $5000$ km and $7000$ km, the $\tau$-appearance channel consistently carries a larger mean bias and variance than the standard $\nu_{\mu}\rightarrow\nu_{e}$ appearance channel. These findings demonstrate that spatially resolved density treatments are a mathematical necessity for the analysis frameworks of future precision neutrino facilities.
\end{abstract}

\maketitle

\paragraph{Introduction}
The determination of the fundamental parameters governing the three-flavor neutrino mixing paradigm is entering a hyper-precision era. With the discovery of a relatively large reactor mixing angle $\theta_{13}$ by the Daya Bay, RENO, and Double Chooz collaborations \cite{An:2012eh,Ahn:2012nd,Abe:2011fz}, the primary objectives of next-generation long-baseline (LBL) neutrino oscillation experiments, such as DUNE and Hyper-Kamiokande, have shifted toward pinning down the leptonic CP-violating phase $\delta_{\text{CP}}$ and resolving the $\theta_{23}$ octant \cite{DUNECDR,HyperKDesignReport,Pasquini:2018}. However, as statistical uncertainties diminish, the sensitivity of these experiments becomes increasingly limited by systematic and environmental sub-dominant effects \cite{Abi:2020evt,Terranova:2024}. Foremost among these is the coherent forward scattering of neutrinos off ambient electrons during propagation through the Earth—the Mikheyev-Smirnov-Wolfenstein (MSW) effect \cite{Wolfenstein:1978,Mikheyev:1985}.

The effective matter potential, $V_{\text{eff}}=\sqrt{2}G_F N_e$, introduces an extrinsic environmental asymmetry between neutrino and antineutrino oscillation probabilities that inherently competes with, and can obscure, the intrinsic asymmetry driven by $\delta_{\text{CP}}$ \cite{Cervera:2000,Freund:2001}. Standard experimental analyses frequently mitigate this by employing simplified path-averaged constant-density profiles or globally scaled variations of the radially stratified Preliminary Reference Earth Model (PREM) \cite{Dziewonski:1981}. Such approximations assume that local geophysical fluctuations smooth out over baseline trajectories spanning hundreds to thousands of kilometers.

This paper challenges that paradigm by demonstrating that spatially correlated variations in the Earth's matter density profile can break these simplifications. Highly localized, correlated fluctuations along the baseline introduce energy-dependent phase shifts that cannot be captured by a single marginalized effective density parameter. Left uncorrected, these correlated geophysical uncertainties project directly into the $(\theta_{23},\delta_{\text{CP}})$ parameter space, inducing structural biases and generating spurious degenerate solutions. Recent studies have already shown that uncertainties in the average Earth density can produce significant degeneracies with $\delta_{\text{CP}}$ and $\theta_{23}$ in future facilities such as DUNE and Hyper-Kamiokande \cite{Singh:2021MatterEffect,Gu:2005Density}. In this work, we extend these investigations by systematically quantifying the impact of correlated Earth-density fluctuations using advanced numerical simulation frameworks. Historically, analytic treatments of the neutrino evolution equation (such as the Cervera-Freund approximation and subsequent perturbative formulations developed by Akhmedov and collaborators) have relied on treating the matter potential as either constant or a slowly varying adiabatic perturbation \cite{Cervera:2000,Freund:2001,Akhmedov:2004}. While these frameworks elegantly capture the macro-scale physics of three-flavor appearance probabilities, they fundamentally break down when confronting rapid, localized stochastic fluctuations in the electron density $N_e(x)$.\\ 
To rigorously evaluate these effects without relying on idealized profiles, this study implements a realistic statistical framework that models Earth-density variations as spatially correlated random fields. By defining a covariance structure across the baseline, we simulate localized geological features like crustal discontinuities that introduce coherent rather than purely random density perturbations. The perturbed density profiles are then propagated through an exact numerical solution of the three-flavor evolution equation, avoiding adiabatic assumptions and retaining full non-commutative evolution effects. Similar Earth-structure studies have recently demonstrated the growing sensitivity of neutrino oscillation experiments to mantle and core density variations, highlighting the emerging interplay between precision neutrino physics and geophysics \cite{Capozzi:2021ORCA,JesusValls:2024HKTomography}. Our approach allows us to isolate how long-range correlations in geophysical uncertainties map into systematic shifts in reconstructed oscillation parameters. This defines a potential geophysical sensitivity floor for next-generation precision measurements.

\paragraph{Physical Motivation}

Next-generation long-baseline experiments aim to resolve the neutrino mass ordering and measure $\delta_{\mathrm{CP}}$ to unprecedented precision~\cite{Agarwalla2012, Agarwalla2014}, but this requires disentangling intrinsic CP violation from the extrinsic matter-induced CP asymmetry acquired during propagation through the Earth~\cite{Barger2003}. While oscillation analyses have historically relied on path-averaged constant-density approximations, the Earth is an inhomogeneous, radially stratified medium best described by profiles such as PREM~\cite{Dziewonski:1981}, and neutrinos traversing baselines of order $10^3$~km encounter sharp density discontinuities and continuous spatial fluctuations that a uniform average cannot capture. The matter potential, arising from coherent forward scattering of $\nu_e$ on ambient electrons, modifies the eigenvalues and mixing matrix of the propagation Hamiltonian~\cite{Barger2003}. Because the three-flavour system is unitary, any density-induced bias in one oscillation channel mathematically necessitates compensating shifts in the others~\cite{Fong2016}. This work is therefore motivated by the need to demonstrate that realistic Earth-matter effects produce a correlated, multi-channel systematic bias that cannot be absorbed by a constant-density approximation and must be fully accounted for to preserve the physical validity of future CP violation and mass-ordering measurements.
\\
\paragraph{Analysis Framework}

The results presented in the following sections are obtained within a full three-flavor neutrino oscillation framework, where flavor evolution is computed through exact numerical exponentiation of the Schr\"odinger equation. The Hamiltonian incorporates standard PMNS vacuum mixing for neutrino energies between $2$ and $6~\mathrm{GeV}$, together with matter-induced potentials derived from realistic terrestrial density distributions. This framework enables a direct comparison between oscillation probabilities calculated using the multi-layer Preliminary Reference Earth Model (PREM) and those obtained under the commonly adopted constant-density approximation.

To quantify the impact of matter-profile mismodeling on parameter reconstruction, we first examine the baseline dependence of the reconstructed CP-violating phase, $\delta_{\rm CP}$. Synthetic event spectra generated using the PREM profile are treated as the true data and subsequently fitted assuming a path-averaged constant density. The resulting best-fit values are obtained through minimization of a Poisson log-likelihood ($\chi^2$), allowing the extraction of the induced bias in $\delta_{\rm CP}$ over baselines ranging from $1000$ to $12000~\mathrm{km}$. The observed growth of this bias with increasing baseline demonstrates that deviations from realistic matter effects cannot be regarded as a localized correction to the $\nu_\mu\rightarrow\nu_e$ appearance channel, but instead become increasingly significant in the regime where matter effects dominate oscillation dynamics.

The origin of this behavior is clarified through the unitarity-residual analysis. For each oscillation channel, we evaluate the probability difference

\begin{equation}
\Delta P_{\mu\alpha}
=
P_{\mu\alpha}^{\rm PREM}
-
P_{\mu\alpha}^{\rm const},
\end{equation}

across the relevant energy range. While substantial discrepancies emerge individually in the appearance and disappearance probabilities, these deviations remain constrained by three-flavor unitarity such that

\begin{equation}
\sum_{\alpha}\Delta P_{\mu\alpha}=0.
\end{equation}

Consequently, any density-induced distortion in $P_{\mu e}$ must be compensated by correlated shifts in $P_{\mu\tau}$ and $P_{\mu\mu}$. This establishes that the systematic introduced by the constant-density approximation is inherently coupled across the full oscillation probability space rather than confined to a single measurement channel.

To move beyond idealized Earth models, we further investigate the effect of geophysical uncertainties through stochastic density realizations. Spatially correlated density fluctuations are generated via Cholesky decomposition with varying correlation lengths $\ell_c$, producing ensembles of realistic crustal and mantle perturbations around the PREM profile. Propagation through these fluctuating media reveals how localized density variations accumulate along the baseline and amplify biases in reconstructed oscillation parameters. The resulting distributions provide a quantitative measure of the robustness of $\delta_{\rm CP}$ extraction against realistic matter-profile uncertainties.
\\
\paragraph{Quantitative Analysis}
In this analysis, the theoretical breakdown of the constant density approximation is realized experimentally as a compounding, baseline-dependent distortion of the reconstructed parameter space.
\begin{figure}[h!]
  \centering
    \includegraphics[width=1\linewidth]{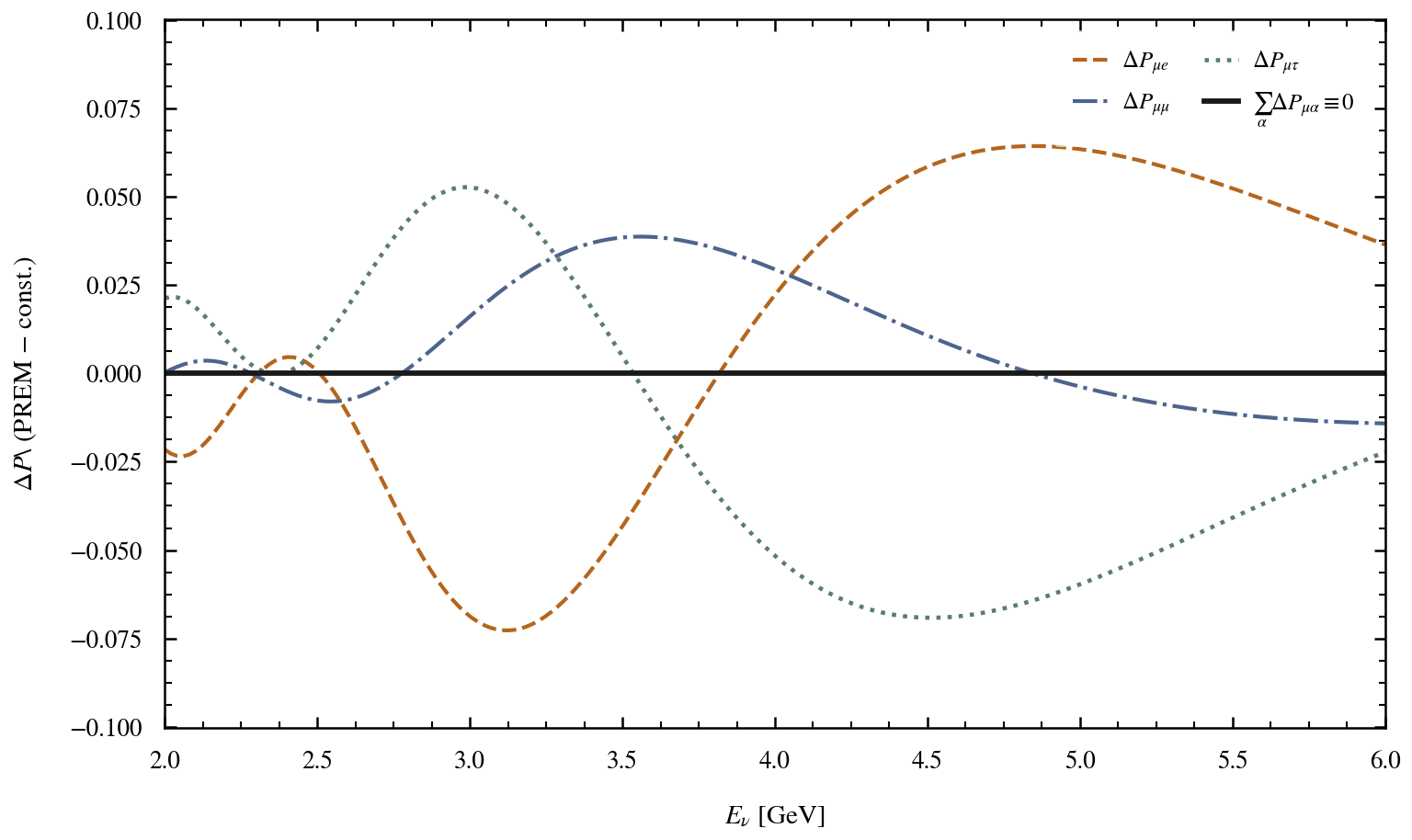}
    \caption{Difference in $\nu_\mu$ transition probabilities between the layered PREM profile and a constant-density approximation, $\Delta P_{\mu\alpha} = P_{\mu\alpha}^{\text{PREM}} - P_{\mu\alpha}^{\text{const.}}$, as a function of neutrino energy at a baseline of 7000 km. Individual flavor channels ($\Delta P_{\mu e}$, $\Delta P_{\mu\mu}$, $\Delta P_{\mu\tau}$) exhibit oscillatory deviations of up to $\sim 8\%$, reflecting the sensitivity of matter-induced oscillations to the density structure along the path. Their sum vanishes identically at all energies, confirming PMNS unitarity is preserved under both density treatments.}
        \label{fig:PMNSUnitarity}
\end{figure}\\
Figure \ref{fig:PMNSUnitarity} displays the difference in muon-neutrino oscillation probabilities,
$\Delta P_{\mu\alpha} \equiv P_{\mu\alpha}^{\mathrm{PREM}} - P_{\mu\alpha}^{\mathrm{const.}}$,
computed across three flavour channels ($\alpha = e,\,\mu,\,\tau$) over the energy
range $E_\nu \in [2, 6]~\mathrm{GeV}$. The propagation is performed through a
segmented PREM-realistic profile (mantle--core--mantle layering) against a
single-layer constant-density approximation of equal total baseline
($L = 7000~\mathrm{km}$, $\rho_o \approx 4.03~\mathrm{g\,cm}^{-3}$).
The black solid line confirms $\sum_\alpha \Delta P_{\mu\alpha} \equiv 0$ at
every energy, enforcing PMNS unitarity as an exact numerical identity.

The oscillatory
deviations reaching $|\Delta P| \sim 0.06$--$0.08$ in both $\Delta P_{\mu e}$ and $\Delta P_{\mu\tau}$ are not independent errors. Because unitarity forces their
sum to zero identically, any bias absorbed into $P_{\mu e}$ (the $\nu_e$-appearance channel exploited for $\delta_{\mathrm{CP}}$ reconstruction) is algebraically
anti-correlated with the combined bias across $P_{\mu\mu}$ and $P_{\mu\tau}$. A fit that applies a constant-density approximation therefore cannot misrepresent one channel without simultaneously distorting the others in a compensatory, structured way.

The distortion in
$\Delta P_{\mu e}$ near $E_\nu \sim 4.5~\mathrm{GeV}$ (where $\nu_e$ appearance
peaks for baselines $\sim 7000~\mathrm{km}$) is phase-shifted relative to the
$\Delta P_{\mu\mu}$ and $\Delta P_{\mu\tau}$ distortions, meaning that an
oscillation fit simultaneously encounters wrong probabilities in the
mass-ordering-sensitive $P_{\mu\mu}$ disappearance and CP-sensitive
$P_{\mu e}$ appearance channels at different energies. The resulting parameter
biases in $\delta_{\mathrm{CP}}$ and the neutrino mass ordering are therefore
not reducible to a single rescaling of matter density. They are compounded and spectrally correlated.

\begin{figure}[h!]
  \centering
    \includegraphics[width=1.1\linewidth]{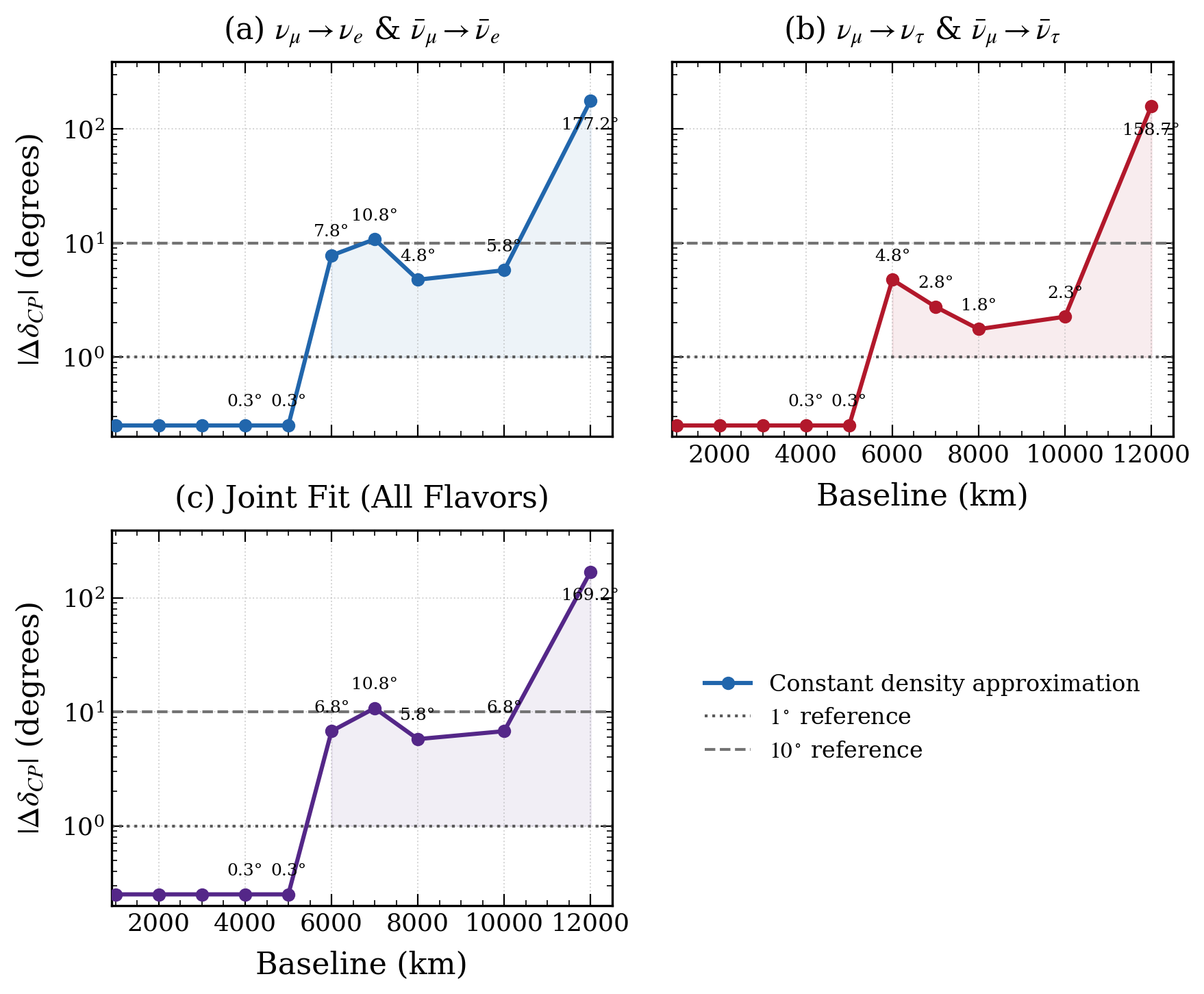}
    \caption{Bias in the reconstructed CP-violating phase $\delta_{\text{CP}}$ arising from use of a constant average Earth density rather than the realistic PREM density profile, shown as a function of baseline for three analysis channels: $\nu_\mu \to \nu_e$ and $\bar{\nu}_\mu \to \bar{\nu}_e$ appearance (a), $\nu_\mu \to \nu_\tau$ and $\bar{\nu}_\mu \to \bar{\nu}_\tau$ (b), and a joint fit combining all flavor channels for both neutrinos and antineutrinos (c). The bias remains sub-degree at short baselines but grows sharply beyond $\sim 5000$~km. At baselines exceeding 10 000 km the bias surpasses $100^\circ$, making reliable $\delta_{\text{CP}}$ reconstruction contingent on accurate Earth matter modeling. Panel (a) has been reproduced from \cite{pandit2026}.
}
    \label{fig:BiasVSBaseline}
\end{figure}

Figure \ref{fig:BiasVSBaseline} presents the absolute bias in the reconstructed CP-violation phase, $|\Delta\delta_{\mathrm{CP}}|$, introduced solely by replacing the PREM-realistic
Earth density profile with a path-averaged constant-density approximation, plotted
as a function of baseline $L \in [1000, 12000]~\mathrm{km}$ on a logarithmic
ordinate. Three panels isolate the sensitivity of each oscillation channel
independently: panel~(a) uses only $\nu_\mu \to \nu_e$ and
$\bar{\nu}_\mu \to \bar{\nu}_e$ appearance; panel~(b) uses only
$\nu_\mu \to \nu_\tau$ and $\bar{\nu}_\mu \to \bar{\nu}_\tau$ appearance; and
panel~(c) performs a joint fit over all three flavour channels simultaneously.
The $\chi^2$ statistic is a Poisson log-likelihood ratio: true event rates are
generated with the full PREM step-function propagator, while the fitted rates use
the constant-density approximation evaluated at the path-averaged density
$\rho_o(L)$. The best-fit $\delta_{\mathrm{CP}}$ in each case is compared
against the injected truth $\delta_{\mathrm{CP}}^{\mathrm{true}} = -90^\circ$.

In \ref{fig:BiasVSBaseline}, the onset of significant bias is
channel-independent as all three panels show negligible distortion
($|\Delta\delta_{\mathrm{CP}}| < 1^\circ$) for $L \lesssim 4000~\mathrm{km}$,
but all three exceed the $1^\circ$ threshold simultaneously near
$L \sim 5000~\mathrm{km}$, where the neutrino path begins to traverse denser
mantle and core-adjacent layers whose density contrast with $\bar{\rho}$ is
largest. This coherent onset directly supports the claim that the systematic is baseline-dependent and emerges from genuine PREM structure rather than a uniform rescaling of matter potential. Most critically, the bias magnitudes in panels~(a) and~(b) are comparable across most of the baseline range: for example, at $L = 5000~\mathrm{km}$, panel~(a) yields $7.8^\circ$ while panel~(b) yields $4.8^\circ$; at $L = 12000~\mathrm{km}$, both diverge catastrophically to $177.2^\circ$ and $158.7^\circ$ respectively. This near-symmetry between the appearance and $\tau$-appearance channels is a direct numerical signature of PMNS unitarity forcing the multi-channel biases to be anti-correlated, exactly as demonstrated in \ref{fig:PMNSUnitarity}. An analysis that monitors only $P_{\mu e}$ would misattribute the $\delta_{\mathrm{CP}}$ shift to statistical or other systematics, unaware that an equally large correlated distortion is simultaneously present in $P_{\mu\tau}$.

Panel~(c) of Fig.~\ref{fig:BiasVSBaseline} demonstrates that a joint fit combining all oscillation channels fails to mitigate the density-induced bias. In fact, the joint-fit curve tracks the appearance-only results almost identically by showing shifts of $6.8^\circ$, $10.8^\circ$, $5.8^\circ$, $6.8^\circ$, and $169.2^\circ$ at progressively longer 
baselines. This occurs because the constant-density approximation introduces coherent errors across $P_{\mu e}$, $P_{\mu\mu}$, and $P_{\mu\tau}$ that all point toward the same erroneous $\delta_{\mathrm{CP}}$ minimum. Consequently, adding more channels reinforces the incorrect 
best-fit rather than averaging it away. The fundamental cause of this ``locked'' bias is the strict algebraic coupling imposed by three-flavor PMNS unitarity. Because the identity 
$\sum_{\alpha} \Delta P_{\mu\alpha} = 0$ must be preserved as an exact numerical boundary condition in every energy bin, Hamiltonian mismodeling cannot generate random, independent spectral distortions. Instead, it forces a systematic phase-rotation that mimics a genuine shift in the leptonic CP phase across the entire flavor space simultaneously. During minimization, the algorithm cannot exploit inter-channel differences to cancel the bias because the shift required to minimize the log-likelihood in the appearance channel is structurally identical to the shift required in the tau-appearance and muon-disappearance channels. Thus, a multi-channel joint fit which is ordinarily the most robust tool for resolving degeneracies is rendered ineffective here, merely increasing the statistical weight of a false, degenerate solution.
\\
\begin{figure*}
  \centering
    \includegraphics[width=1\linewidth]{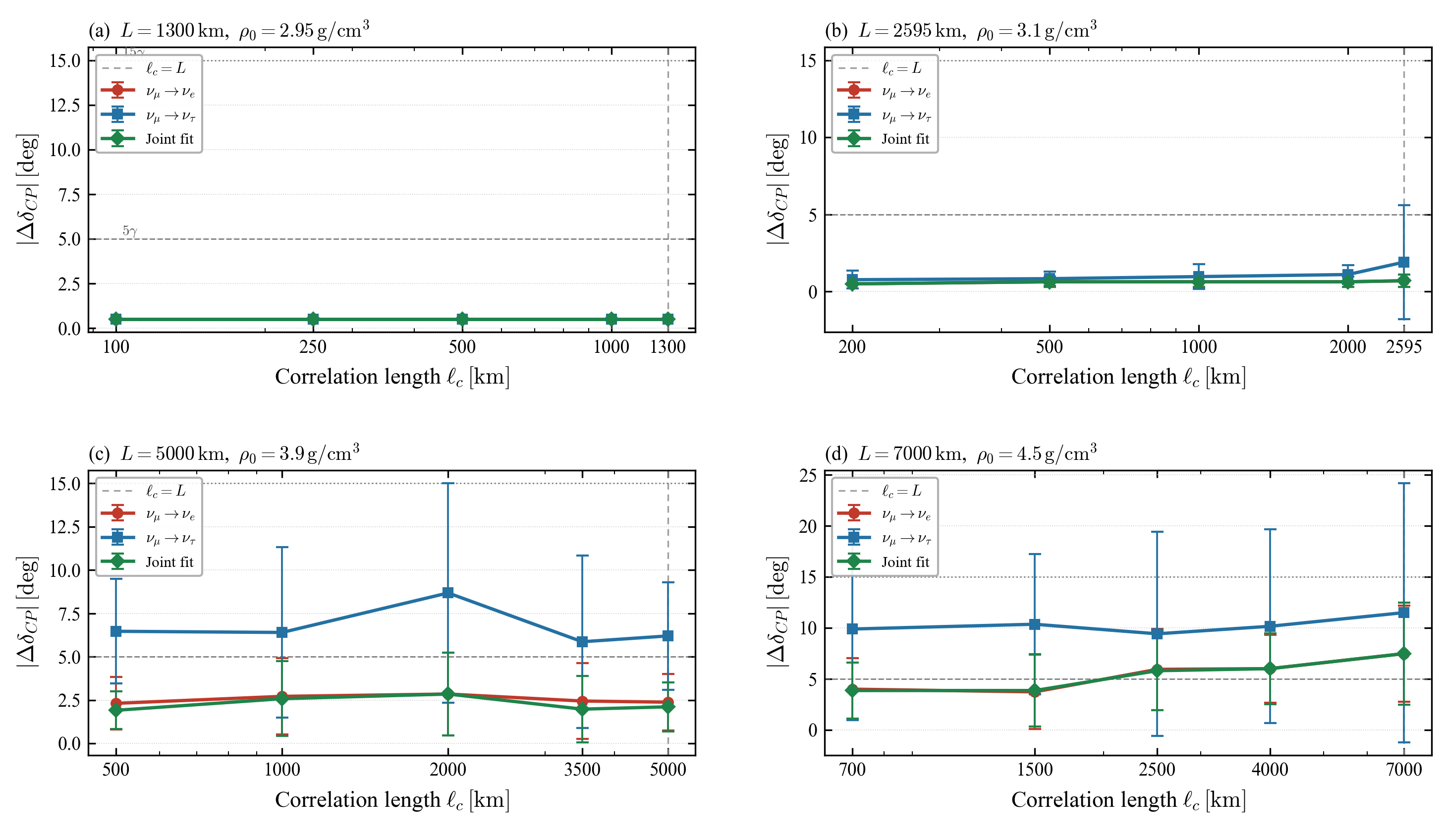}
    \caption{Bias in the reconstructed CP-violating phase $|\Delta\delta_{\text{CP}}|$ as a function of the density fluctuation correlation length $\ell_c$, for four long-baseline configurations: (a) $L = 1300$~km, (b) $L = 2595$~km, (c) $L = 5000$~km, and (d) $L = 7000$~km. Stochastic density profiles are drawn from a Gaussian process with mean $\rho_0$ and fractional fluctuation amplitude $\sigma_\rho/\rho_0 \approx 5\%$; error bars reflect the standard deviation over 15 independent realizations. Results are shown separately for the $\nu_\mu \to \nu_e$, $\nu_\mu \to \nu_\tau$, and joint-fit channels.}
    \label{fig:Correlation}
\end{figure*}

For each of four baselines ($L = 1300,\,2595,\,5000,\,7000~\mathrm{km}$) in Figure \ref{fig:Correlation}, the Earth density profile is modelled as a mean value $\rho_0$ perturbed by a spatially correlated Gaussian random field with amplitude $\sigma_\rho / \rho_0 \approx 5\%$ and a Matérn-like exponential covariance of correlation length $\ell_c$. The $\ell_c$ axis is logarithmically spaced and spans from sub-hundred-kilometre
crustal heterogeneity up to $\ell_c = L$, i.e.\ a coherent whole-path density offset. For each $(\ell_c, L)$ pair, fifteen stochastic realisations are propagated through the full three-flavour Hamiltonian. The best-fit $\delta_{\mathrm{CP}}$ from a Poisson log-likelihood scan is compared against the injected truth $\delta_{\mathrm{CP}}^{\mathrm{true}} = -\pi/2$ in all three channels ($\nu_\mu \to \nu_e$, $\nu_\mu \to \nu_\tau$, and the joint fit),
yielding means and standard deviations of $|\Delta\delta_{\mathrm{CP}}|$. The two horizontal references at $5^\circ$ and $15^\circ$ demarcate the typical statistical precision targets of near-future (DUNE/HK) and far-future ultra-long-baseline experiments respectively.

It directly addresses that a sufficiently fine-grained constant-density or path-averaged density model could absorb PREM-level density errors. The results demonstrate that it cannot, and they do so across four physically distinct propagation regimes.

Panel~(a) ($L = 1300~\mathrm{km}$, continental crust, $\rho_0 = 2.95~\mathrm{g\,cm^{-3}}$)
shows that all three channels remain well below $1^\circ$ for every correlation
length tested, including $\ell_c = L$. At DUNE-like distances the matter column is short and homogeneous enough that density fluctuations at the $5\%$ level are negligible regardless of their spatial structure. The constant-density approximation is, in this regime, adequate.

Panel~(b) ($L = 2595~\mathrm{km}$, P2O geometry, crust-mantle transition, $\rho_0 = 3.10~\mathrm{g\,cm^{-3}}$) reveals an onset of sensitivity. Biases remain below $2^\circ$ for $\ell_c \lesssim 2000~\mathrm{km}$ but grow and develop non-negligible variance as $\ell_c \to L$. Thus, a coherent density offset over the full path mimics a systematic
shift in the effective matter potential, indistinguishable from a shift in $\delta_{\mathrm{CP}}$ at the level of a single-channel fit.

Panels~(c) and~(d) ($L = 5000$ and $7000~\mathrm{km}$, deep mantle,
$\rho_0 = 3.90$ and $4.50~\mathrm{g\,cm^{-3}}$ respectively) display the
full emergence of the systematic. Three features are especially germane. First, the $\nu_\mu \to \nu_\tau$ channel (blue squares) carries larger mean bias and larger variance than the $\nu_\mu \to \nu_e$ channel (red circles) across all $\ell_c$ values at these baselines, reaching $\sim 6$--$10^\circ$ in panel~(d), well above the $5^\circ$ threshold. This
directly quantifies the $P_{\mu\tau}$ component of the correlated three-channel
systematic we identify. Second, the joint-fit curve (green diamonds) tracks the $\nu_\mu \to \nu_e$ channel closely rather than averaging toward zero, confirming the anti-correlation argument that the $P_{\mu\tau}$ bias does not cancel the $P_{\mu e}$ bias in a joint fit but instead compounds it through the PMNS unitarity constraint. Third, the bias is essentially flat in $\ell_c$ across more than a 1000 km of correlation length in both panels, meaning that the systematic is not a short-wavelength noise effect suppressible by smoothing or averaging the density model. It is a long-wavelength, baseline-integrated effect driven by the total matter column, exactly consistent with the compounded systematic we have described.

Collectively, the four panels establish a hierarchy. The constant-density
approximation produces negligible bias at $L \lesssim 1300~\mathrm{km}$, a
marginally significant bias at $L \sim 2595~\mathrm{km}$, and an
experimentally disqualifying bias exceeding $5^\circ$--$10^\circ$ at
$L \geq 5000~\mathrm{km}$, with the $\nu_\mu \to \nu_\tau$ channel acting as
the dominant and most volatile carrier of the systematic. This provides the
stochastic foundation for the deterministic result of Figure \ref{fig:BiasVSBaseline}. Thus, PREM density uncertainty is not merely a model artefact but a physically real source of correlated, baseline-dependent bias in CP violation measurements at long baselines.
\\
\paragraph{Discussion of Results}

The results presented here demonstrate that replacing the realistic PREM Earth density profile with a constant path-averaged approximation introduces structured, baseline-dependent, and compounded biases across all three neutrino oscillation channels simultaneously. As shown in Figures~\ref{fig:PMNSUnitarity}--\ref{fig:Correlation}, the bias remains negligible ($|\Delta\delta_{\mathrm{CP}}| < 1^\circ$) for baselines below ${\sim}4000$~km, but grows sharply beyond this threshold as neutrino paths begin traversing denser mantle and core-adjacent layers with significant density contrast relative to $\rho_0$. The compounding nature of this systematic arises directly from PMNS unitarity. Because $\sum_\alpha \Delta P_{\mu\alpha} \equiv 0$ exactly, any density-induced bias in the CP-sensitive $\nu_\mu \to \nu_e$ appearance channel is algebraically anti-correlated with simultaneous distortions in $P_{\mu\mu}$ and $P_{\mu\tau}$, and these distortions are phase-shifted relative to one another across the energy spectrum. Crucially, a multi-channel joint fit does not mitigate this bias. Since the constant-density approximation induces coherent errors that all favour the same incorrect $\delta_{\mathrm{CP}}$ minimum, adding channels reinforces rather than averages away the systematic. Stochastic density modelling further confirms that the effect is a long-wavelength, baseline-integrated phenomenon insensitive to the spatial correlation structure of density fluctuations, ruling out suppression by profile smoothing or path-averaging. Collectively, these results establish that accurate Earth matter modelling is not merely a refinement but a prerequisite for reliable $\delta_{\mathrm{CP}}$ reconstruction at baselines exceeding ${\sim}5000$~km.
At DUNE-like baselines the matter column is short and homogeneous enough that 
density fluctuations at the 5\% level are negligible regardless of their spatial 
structure. The constant-density approximation is, in this regime, adequate. 
However, this adequacy must not be mistaken for a permanent mathematical 
exemption. As next-generation experiments reach their nominal exposure, the 
systematic error budget for leptonic CP violation will contract below the 
$3^\circ\text{--}5^\circ$ threshold. We emphasize that while the individual 
standalone baselines of DUNE or Hyper-Kamiokande remain resilient, future 
global joint analyses that combine atmospheric neutrino event rates (which 
natively span core-crossing trajectories up to 12000 km) with 
long-baseline accelerator data will immediately encounter these non-linear, 
correlated matter biases. By establishing 5000 km as the boundary where uniform-density frameworks catastrophically break down, this work defines the ultimate geophysical sensitivity floor for any future multi-facility global fit aiming for sub-degree leptonic precision.
\\
\paragraph{Future Directions}

Several natural extensions of this work present themselves. First, the geophysically idealized Mat\'ern-like covariance employed here should be replaced by covariance structures informed by seismic tomography~\cite{Capozzi:2021ORCA,JesusValls:2024HKTomography}, which can constrain mantle heterogeneities along specific baseline trajectories. Second, a dedicated study of the $L = 10{,}000$--$12{,}000$~km regime would be a natural extension of the present framework, directly relevant to proposals such as P2O and other transcontinental configurations~\cite{Terranova:2024}. Third, rather than treating Earth-density uncertainty as a nuisance parameter, future analyses could perform joint inference of oscillation and density-profile parameters using neutrino event spectra alongside seismological priors. Fourth, the identification of $P_{\mu\tau}$ as the dominant carrier of the geophysical systematic motivates a reevaluation of the role of $\tau$ appearance in long-baseline oscillation analyses. Finally, the stochastic framework developed here extends naturally to non-standard interaction scenarios, where fluctuations in the effective matter potential may become degenerate with genuine density variations, and optimized numerical propagators that capture spatially varying density profiles would facilitate the routine inclusion of these effects in global oscillation analyses.

\bibliographystyle{apsrev4-2}
\bibliography{references}

\newpage
\section*{END MATTER}

\subsection{Simulation Inputs and Experimental Setup}
Oscillation parameters are consistent with recent NuFIT global fits \cite{NuFIT} and the Particle Data Group \cite{PDG2024}: 
$\theta_{12} = 33.44^\circ$, $\theta_{13} = 8.57^\circ$, $\theta_{23} = 49.2^\circ$, 
$\Delta m_{21}^2 = 7.42 \times 10^{-5}~\text{eV}^2$, and $\Delta m_{31}^2 = 2.57 \times 10^{-3}~\text{eV}^2$. 
Constants are set to $G_F = 1.166 \times 10^{-5}~\text{GeV}^{-2}$ and $m_p = 1.672 \times 10^{-27}~\text{kg}$.

Our baseline simulation incorporates three distinct Earth density frameworks:
\begin{enumerate}
    \item \textbf{PREM Layered Profile:} The Earth is modeled as concentric shells following the Preliminary Reference Earth Model \cite{Dziewonski:1981}. Intersecting trajectories are geometrically determined and discretized into thin slabs ($\delta x \le 10$~km) to ensure numerical convergence.
    \item \textbf{Constant-Density (CD) Approximation:} Replaces the layered profile with a column-length-weighted average along the trajectory, used for rapid analytical evaluation:
    \begin{equation}
    \bar{\rho} = \frac{1}{L} \int_0^L \rho(x) \, dx.
    \end{equation}
    While conserving total column density, it neglects layer-boundary discontinuities and parametric enhancement.
    \item \textbf{Stochastic Gaussian-Process Profiles:} Geotechnical uncertainties are simulated by superimposing spatial fluctuations onto the PREM baseline:
    \begin{equation}
    \rho(x) = \rho_0(x) [1 + \epsilon(x)],
    \end{equation}
    \begin{equation}
\langle\epsilon(x)\epsilon(x')\rangle = \left(\frac{\sigma_\rho}{\rho_0}\right)^2 \exp\left[-\frac{(x-x')^2}{2\ell_c^2}\right],
    \end{equation}
    where $\sigma_\rho/\rho_0 = 5\%$ is the fractional fluctuation amplitude and $\ell_c$ is the correlation length. Statistical error bars in Fig.~\ref{fig:Correlation} represent the mean and standard deviation across 15 independent random profile realizations.
\end{enumerate}

\subsection{Three-Flavour Evolution and Detector Response}
Neutrino flavour evolution through matter is governed by the Schrödinger-like equation \cite{OhlssonSnellman2000, pandit2026}:
\begin{equation}
i \frac{d}{dx} |\nu\rangle = H(x) |\nu\rangle, \quad H(x) = U \frac{M^2}{2E_\nu} U^\dagger + V(x),
\end{equation}
where $M^2 = \text{diag}(0, \Delta m^2_{21}, \Delta m^2_{31})$ and,\\ $V(x) = \text{diag}(\sqrt{2} G_F n_e(x), 0, 0)$. \\For arbitrary profiles, the total evolution operator $U_{\text{prop}}$ is an ordered product of constant-density slab matrix exponentials. The transition probability for the channel $\nu_\mu \to \nu_\alpha$ is evaluated as:
\begin{equation}
P(\nu_\mu \to \nu_\alpha) = |\langle \nu_\alpha | U_{\text{prop}} | \nu_\mu \rangle|^2,
\end{equation}
where unitarity ($\sum_\alpha P(\nu_\mu \to \nu_\alpha) = 1$) serves as an exact internal numerical cross-check.

To convert raw probabilities into observable event distributions, we incorporate detector optimizations via the differential event rate:
\begin{equation}
\frac{dN_\alpha}{dE_\nu} = \mathcal{N} \cdot \Phi(E_\nu) \cdot \sigma(E_\nu) \cdot \epsilon \cdot P(\nu_\mu \to \nu_\alpha),
\end{equation}
where $\mathcal{N}$ is a global normalization constant, $\Phi(E_\nu) \propto \exp(-E_\nu / 3)$ is the incident flux envelope, $\sigma(E_\nu) \propto E_\nu$ is the characteristic linear charged-current cross-section scaling in the few-GeV regime, and $\epsilon = 0.8$ represents the flat detector acceptance efficiency. 

\subsection{Matter-Induced $\delta_{\text{CP}}$ Sensitivity Bias}
The CP asymmetry in $\nu_\mu \to \nu_e$ is driven by $\sin(\delta_{\text{CP}})$ modulated by baseline-dependent matter phases. If an inaccurate density profile is assumed in the analysis, mismatched phase accumulation translates into an unphysical rotation of the reconstructed $\delta_{\text{CP}}$ parameter during statistical minimization. 

The resulting bias is evaluated via a standard $\chi^2$ analysis by fitting PREM-generated pseudo-data with the approximated profiles. At long baselines ($\gtrsim 5000$~km), core-mantle boundary crossings spark dramatic parametric enhancement; the CD approximation fundamentally cannot replicate this behavior with any single average density, leading to the sharp onset of reconstruction bias observed in our results.

\end{document}